\documentclass[lettersize,journal]{IEEEtran}
\usepackage{amsmath,amsfonts}
\usepackage{algorithmic}
\usepackage{algorithm}
\usepackage{array}
\usepackage{multirow}
\usepackage[caption=false,font=normalsize,labelfont=sf,textfont=sf]{subfig}
\usepackage{textcomp}
\usepackage{stfloats}
\usepackage{url}
\usepackage{verbatim}
\usepackage{graphicx}
\usepackage{cite}
\usepackage[hidelinks]{hyperref}
\usepackage{xcolor}

\newcommand{\revision}[1]{{\color{black}#1}}

\begin{document}

\title{Collective Communication for Distributed LLM Systems: Planning, Runtime Adaptation, and Computation Coordination}

\author{Xuebin Song, Menghao Zhang, Yuezheng Liu, Jinyi Xia, Shucan Yang, Xiaohe Hu, Chunming Hu, Mingwei Xu
\thanks{Xuebin Song, Menghao Zhang, Yuezheng Liu, Jinyi Xia, Shucan Yang and Chunming Hu are with the School of Software, Beihang University, Beijing, China. Xuebin Song, Menghao Zhang and Mingwei Xu are also with the State Key Laboratory of Internet Architecture, Tsinghua University, Beijing, China. Xiaohe Hu is with the Infrawaves, Beijing, China. Menghao Zhang is the corresponding author.}
}

\markboth{IEEE Network}%
{Shell \MakeLowercase{\textit{Song et al.}}: Collective Communication for Distributed LLM Systems: Planning, Runtime Adaptation, and Computation Coordination}


\maketitle

\begin{abstract}
Distributed large language model (LLM) systems increasingly rely on collective communication primitives such as AllReduce (AR), ReduceScatter (RS), AllGather (AG), and AlltoAll (A2A). In modern LLM training and serving clusters, heterogeneous GPU interconnects, multi-NIC networking, mixed parallelism strategies, low-latency inference requests, and high-throughput training pipelines have motivated increasingly diverse ways to plan, execute, and overlap collective communication. This paper presents a tutorial-style, collective-centric taxonomy for collective communication. We organize recent advances into three layers:
communication planning, which generates topology-aware collective schedules; communication execution and adaptation, which maps these schedules onto GPU runtimes and hardware in real clusters; and computation-communication coordination, which turns collective optimization into end-to-end training and inference benefits. We further discuss open challenges and future opportunities for collective communication in distributed LLM systems.
\end{abstract}


\section{Introduction}

\IEEEPARstart{L}{arge} language models (LLMs) have advanced rapidly in recent years, reshaping how intelligent services are built and deployed across many application domains.  
This progress is driven by the continual growth of model parameters, training corpora, context lengths, and serving demands. A single GPU is often insufficient for either storing model states or providing required throughput. Modern LLM systems therefore scale out across many GPUs and often across many servers. This scale-out step is necessary, but it is not free. Once a model is partitioned across devices, the devices must repeatedly exchange activations, gradients, parameters, or intermediate results. These communication phases can keep expensive accelerators idle: On a DGX-H100 cluster connected by 400 GB/s InfiniBand, when training GPT-3 and Llama-2 with tensor parallelism (TP), collective communication accounts for 22\%$\sim$47\% of end-to-end iteration time;
for GPT-3-13B specifically, this fraction ranges from 17\% to 43\% when scaling from 8 to 32 H100 GPUs~\cite{domino}. Communication optimization is therefore a prerequisite for efficient large-scale LLM training and serving.

The communication pressure arises from the parallelization mechanisms that make LLMs feasible in the first place. A model can be replicated across workers, split by layers, partitioned within a layer, or routed through different experts. Each choice creates a different communication pattern. Data parallelism (DP) synchronizes model updates across replicas; TP exchanges partial results within model layers; pipeline parallelism (PP) transfers intermediate activations across stages; and expert parallelism (EP) routes tokens among distributed experts. These mechanisms rely heavily on collective communication primitives, including AR, RS, AG, and A2A. Consequently, collective communication is no longer only a hidden library routine underneath distributed training frameworks. It has become a common systems substrate whose efficiency affects training throughput, inference latency, and service-level objectives (SLOs).

The importance of collectives has motivated a broad range of optimization efforts. For example, Themis~\cite{themis} improves collective communication under heterogeneous bandwidth by scheduling data chunks according to the capacity of different network dimensions. AutoCCL~\cite{autoccl} shows that collective performance is sensitive to runtime configuration and automatically tunes communication parameters for different models, messages, and system conditions. CoCoNet~\cite{coconet} optimizes communication together with computation by transforming programs to split, reorder, fuse, and overlap communication operations. These examples suggest that collective communication optimization has moved beyond selecting a fixed ring, tree, or hierarchical algorithm; it now involves schedule design, runtime configuration, hardware resource utilization, and computation-communication coordination.

Recent work on distributed-training communication provides an architecture-oriented view by
organizing optimization around parallelization strategies, collective communication libraries (CCLs),
and networks~\cite{wei2024communication}. \revision{In contrast, our survey takes a complementary collective-centric view. It follows a collective operation through three stages: communication planning, communication execution and adaptation, and computation-communication coordination. This perspective connects systems that optimize different parts of a collective's lifecycle and relates their effects to end-to-end LLM training and serving performance.}

Rather than providing an exhaustive bibliography, this paper offers a tutorial map of
collective communication optimization for distributed LLM systems. We organize recent advances into
a three-layer system view. Communication planning determines the logical schedule of a collective
operation according to the primitive type, workload, and cluster topology.
Communication execution and adaptation translate this schedule into efficient behavior on real GPU
clusters by considering runtime abstractions and hardware resources. Computation-communication coordination then integrates collective
communication with model execution so that communication optimization can improve end-to-end
training and inference performance. Together, these three layers describe the path from collective algorithm design to practical system benefit.
\revision{Note that the three-layer taxonomy is intended to organize the literature, rather than to impose strict boundaries between systems. In practice, individual systems can involve techniques at more than one layer. We classify each work according to its primary optimization decision and main integration point. This convention provides a structured view of a field with overlapping techniques. Cross-layer mechanisms are discussed where they are important to understanding a system.}

The main contributions of this paper are summarized as follows:
\begin{enumerate}
    \item We introduce the collective primitives, parallelism patterns, and workload metrics needed
    to understand collective communication in distributed LLM systems.
    \item We present a collective-centric system view that connects communication planning,
    communication execution and adaptation, and computation-communication coordination.
    \item We review representative advances in each layer and emphasize the evolution of research challenges in this domain rather than presenting a flat list of techniques.
    \item We discuss open challenges and future opportunities for collective communication in
    distributed LLM systems.
\end{enumerate}

The rest of this paper is organized as follows. Section~\ref{sec:background} introduces the background, including collective primitives, LLM parallelism patterns, and optimization metrics. Section~\ref{sec:overview} explains the system view used to organize the paper. Sections~\ref{sec:comm plan}, \ref{sec:comm exec}, and \ref{sec:codesin} then describe communication planning, communication execution and adaptation, and computation-communication coordination, respectively. Section~\ref{sec:challenges} discusses open challenges and future opportunities, and Section~\ref{sec:conclusion} concludes this paper.

\section{Background}
\label{sec:background}
\begin{figure*}[!t]
\centering
\includegraphics[width=\textwidth]{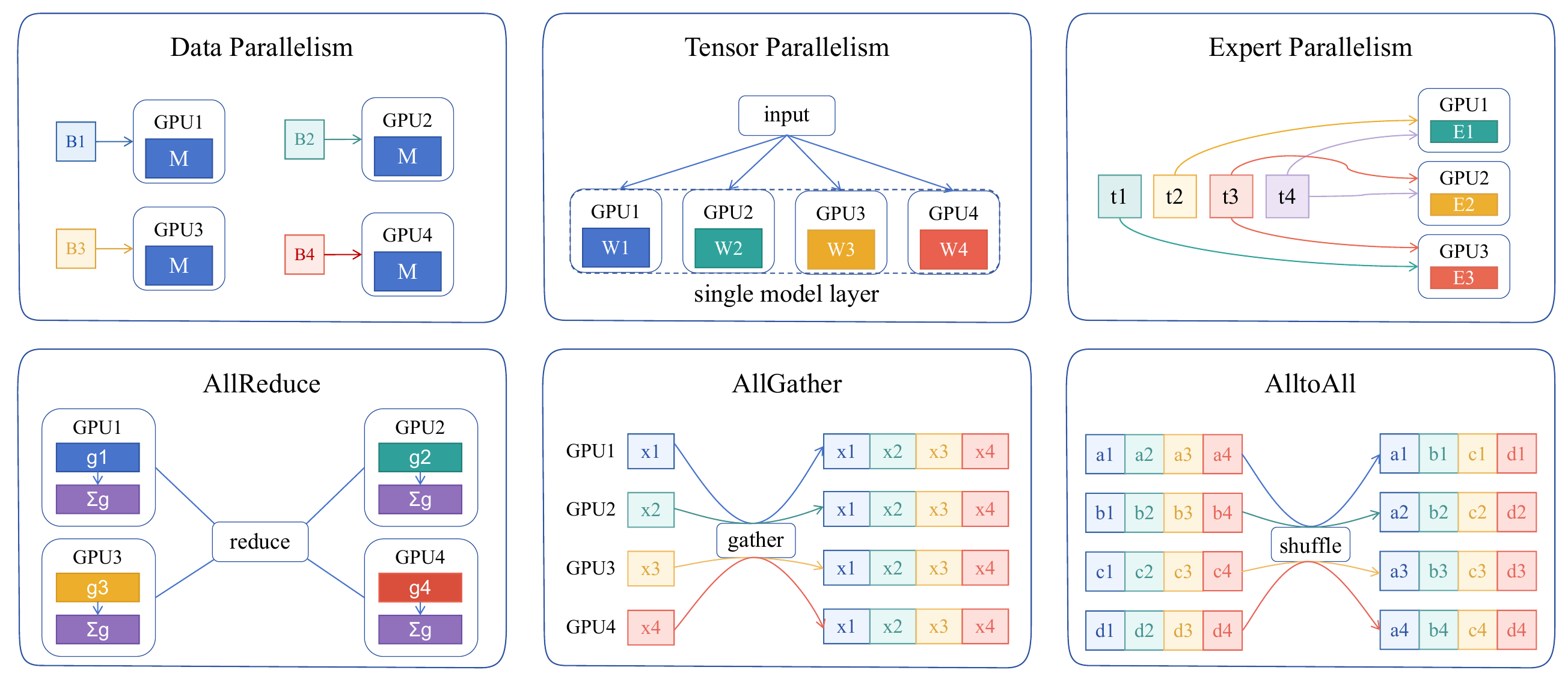}
\caption{Illustration of representative LLM parallelism mechanisms (DP, TP, and EP) and common collective communication primitives (AR, AG, and A2A). \revision{RS is not included in the figure because of layout constraints.}}
\label{fig_parallelism_collectives}
\end{figure*}

This section introduces the background needed to understand collective communication
optimization in distributed LLM systems. We begin with collective communication primitives and then
trace how LLM training and serving workloads instantiate them under different parallelism and
service scenarios. Fig.~\ref{fig_parallelism_collectives} illustrates representative LLM parallelism mechanisms and common collective communication primitives introduced in this section.

\subsection{Collective Communication Primitives}
Collective communication refers to structured data movement among a group of participating
workers. Compared with point-to-point communication, a collective primitive describes a recurring
traffic pattern over the entire worker group and is usually implemented by a collective
communication library. The same primitive can have different algorithmic realizations, such
as ring, tree, hierarchical, topology-aware, or synthesized schedules. For example, two
systems may both call AR, but one implementation may send data around a logical ring while
another may use a tree or a hierarchy that follows the physical cluster topology. Therefore, a
primitive name describes the communication semantics, while the schedule determines how the network
is actually used.

AR aggregates values from all workers and returns the reduced result to every worker.
It is widely used for synchronizing gradients or partial results. Broadcast sends one
worker's data to the rest of the group, which is useful for initialization or parameter
distribution. RS first reduces distributed chunks and leaves each worker with one shard
of the reduced result, while AG performs the complementary operation by collecting shards
from all workers. These two primitives are important in sharded training and TP
execution because they reduce redundant data movement and memory pressure. A2A exchanges
different chunks among all participants. It is particularly important for mixture-of-experts (MoE)
models, where tokens are dispatched to different experts and later combined. Point-to-point transfers also appear in pipeline execution, but
AR, RS, AG, and A2A form the core collective patterns considered in
this paper.

\subsection{Collective Demands in Distributed LLM Systems}
LLM systems create collective communication demand through the parallelization mechanisms
used to scale model execution. A useful way to understand these mechanisms is to ask what is
replicated, what is partitioned, and what must be exchanged after the partitioning. DP replicates
the model across workers and therefore requires gradient or parameter
synchronization, commonly implemented through AR or through RS and AG in
sharded variants. It is conceptually simple, but its synchronization traffic grows with model
size and the number of replicas. TP partitions operators within a model layer,
which introduces additional AR operations to aggregate partial operator results. It
allows several GPUs to compute different parts of the same layer, so the partial results must be
exchanged before the layer can produce a complete output. PP divides model layers across stages; although its basic communication is
often between adjacent stages, it interacts with DP and TP to form more complex
communication graphs. EP, which is widely used in MoE models, introduces dynamic
A2A communication because different tokens are routed to different experts. Unlike
the repeated synchronization in DP, MoE communication depends on the token routing
decision and can vary across batches.

Serving workloads further change the role of collectives. Training often executes repeated
communication patterns over relatively stable iteration graphs, whereas inference workloads are
driven by request length, batch composition, routing decisions, and service-level objectives.
MoE serving makes dispatch and combine operations latency-sensitive.
Small batches, bursty arrivals, and changing expert
popularity can make the same collective primitive behave differently from one serving interval to
the next. As a result, collective communication in distributed LLM systems should be understood
not only as a training-time synchronization mechanism, but also as a serving-time data movement
substrate for latency-sensitive routing and aggregation.

\subsection{Optimization Objectives and Metrics}
The objective of collective communication optimization depends on where the primitive appears
in the workload. In training, the primary metrics include iteration time, throughput, scaling
efficiency, exposed communication time, and achieved bandwidth. A faster collective is useful only
when it reduces the communication time that remains on the critical path of the training iteration.
This is why many systems split, reorder, or overlap collectives with computation rather than only
improving microbenchmark bandwidth.

Inference introduces a different set of metrics. Time to first token (TTFT), time per output
token (TPOT), tail latency, request priority, and service-level objectives become central.
A schedule that performs well for large, repeated training messages may not be suitable for
small-batch decoding, bursty expert routing, or multi-tenant serving environments. System-level
metrics such as GPU idle time, NIC utilization, multi-NIC balance, runtime overhead, and
predictability also matter because they determine whether communication optimization translates into
end-to-end service performance. These metrics also explain why this paper does not treat
collective optimization as a single library problem. 
The best schedule may fail if the runtime maps it poorly to GPU and NIC resources, and a faster primitive may not help if it is hidden
outside the workload's critical path. These distinctions motivate the taxonomy in the next section,
which follows collective optimization from schedule planning, to execution and adaptation, and
finally to computation-communication coordination.

\section{System Overview}
\label{sec:overview}

\begin{figure}[!t]
\centering
\includegraphics[width=\linewidth]{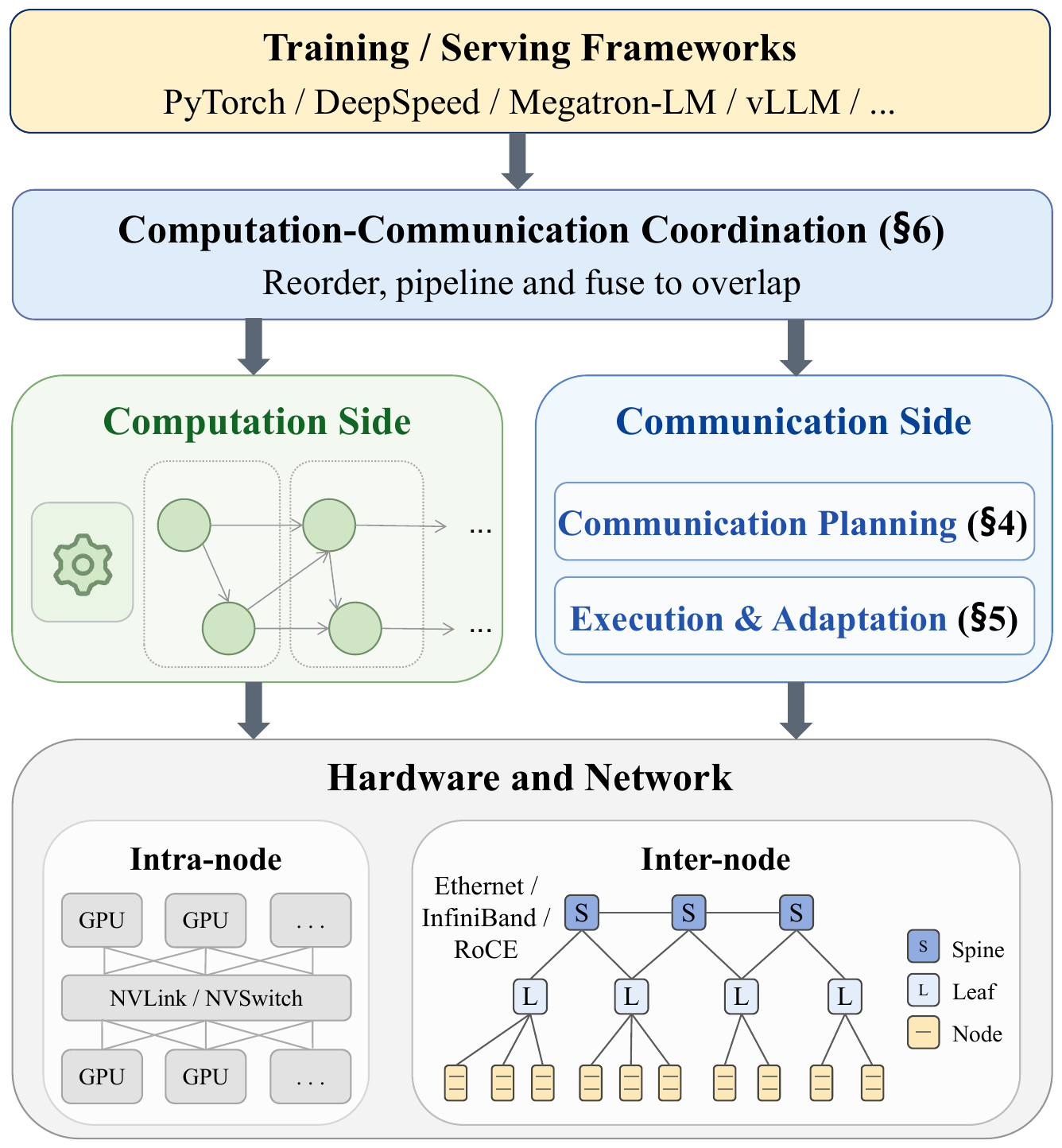}
\caption{System view of collective communication
optimizations in distributed LLM systems.}
\label{fig_system_framework}
\end{figure}

Fig.~\ref{fig_system_framework} presents the system view used to organize this paper. It
abstracts distributed LLM systems from the perspective of collective communication optimization,
rather than describing a specific deployment. The figure connects workload demands, collective
communication mechanisms, hardware resources, and end-to-end metrics. This view emphasizes that
collective communication is shaped jointly by parallelism strategy, serving behavior, communication runtimes, and cluster interconnects.

Workload demand is the entry point of the framework. Training and serving frameworks choose
parallelism strategies, batch structures, model partitions, and routing policies. These choices
determine the type, frequency, message size, and latency sensitivity of collective operations.
Gradient synchronization in DP training creates repeated AR traffic, whereas
expert routing in MoE inference creates token-dependent A2A traffic. Thus, a collective
operation should be interpreted together with the workload context in which it appears.

The communication side of the framework contains two closely related stages. Communication
planning designs the logical schedule of a collective, including the participating workers, data
chunking strategy, communication order, and topology assumptions. Communication execution and
adaptation then realize this logical schedule on a deployed cluster. This realization depends on GPU
kernels, channels, streams, NIC binding, transport protocols, path quality, and runtime
configuration. A plan that is efficient at the algorithmic level may still underperform if it is not
matched to the available GPU and network resources.

The computation side determines whether an optimized collective improves end-to-end
performance. A faster collective primitive does not necessarily reduce training iteration time or
serving latency if the communication is already hidden or if another operator dominates the critical
path. Conversely, even a modest collective improvement can be important when communication blocks
useful computation. Computation-communication coordination therefore forms a separate part of the
system view, connecting collective communication with model graphs, operator dependencies, GPU
resource contention, and serving schedules.

Based on this system view, the paper organizes collective communication optimization into
three categories. Section~IV discusses communication planning, which designs collective schedules
under workload and topology constraints. Section~V discusses communication execution and adaptation,
which maps these schedules onto real GPU clusters and network conditions. Section~VI discusses
computation-communication coordination, which determines whether collective optimization translates
into end-to-end benefits for training and inference workloads.
Table~\ref{tab:taxonomy-comparison} summarizes representative systems across
the three layers of the taxonomy. Integration scope reflects which part of the
software stack a system modifies. Workload distinguishes training-oriented
designs from inference-oriented and general communication-layer approaches.

\revision{This paper aims to provide a tutorial map rather than an exhaustive bibliography. We review recent work from leading systems, networking, and ML-systems venues, with a focus on collective communication optimization. Guided by the three-layer framework, we select representative and influential work from each direction to explain its core ideas and technical evolution. Given the space constraints of this survey, we select only a few representative works for each direction.}

\revision{Several related directions also affect collective communication performance but are outside the main scope of this survey. In-network aggregation moves part of the reduction into network devices and mainly concerns network data-plane capabilities and deployment protocols. TP overlap is only one part of the broader computation-communication overlap literature. This survey takes a collective-centric view of communication optimization itself and discusses only representative work on computation-communication overlap.}

\begin{table*}[!t]
\caption{Comparison of Representative Collective Communication Optimization Systems}
\label{tab:taxonomy-comparison}
\centering
\scriptsize
\setlength{\tabcolsep}{1.7pt}
\renewcommand{\arraystretch}{1.12}
\begin{tabular}{|
>{\raggedright\arraybackslash}m{1.70cm}|
>{\raggedright\arraybackslash}m{1.25cm}|
>{\raggedright\arraybackslash}m{4.00cm}|
>{\raggedright\arraybackslash}m{3.50cm}|
>{\raggedright\arraybackslash}m{1.45cm}|
>{\raggedright\arraybackslash}m{2.00cm}|
>{\raggedright\arraybackslash}m{1.05cm}|
}
\hline
\textbf{Work} &
\textbf{Layer} &
\textbf{Core mechanism} &
\revision{\textbf{Performance}} &
\textbf{Decision stage} &
\textbf{Integration scope} &
\textbf{Workload} \\
\hline
BlueConnect~\cite{blueconnect} &
\multirow[c]{4}{=}[-2.8\baselineskip]{Planning} &
Decomposes AR into hierarchical RS and AG phases to reduce slow-link traffic &
\revision{87\% sync-overhead reduction over GLOO} &
Offline &
Collective operator &
Training \\
\cline{1-1}\cline{3-7}
Themis~\cite{themis} &
&
Schedules chunks across heterogeneous links to balance network dimensions &
\revision{1.25--1.49$\times$ training speedup} &
Runtime &
CCL scheduler &
Training \\
\cline{1-1}\cline{3-7}
SCCL~\cite{sccl} &
&
Synthesizes topology-specific collective schedules with constraint solving &
\revision{Up to 2.2$\times$ AG speedup over NCCL} &
Offline &
Generated schedule &
General \\
\cline{1-1}\cline{3-7}
TE-CCL~\cite{teccl} &
&
Uses multi-commodity flow to route and schedule chunks under link capacities &
\revision{Up to 3.18$\times$ algorithm-bandwidth improvement over RCCL} &
Offline &
Solver-generated schedule &
General \\
\hline
MSCCL++~\cite{mscclpp} &
\multirow[c]{3}{=}[-2.2\baselineskip]{Execution} &
Exposes GPU-side primitives and channel abstractions for programmable collectives &
\revision{1.7$\times$ communication speedup and 1.2$\times$ inference speedup} &
Program./\allowbreak compile &
Communication stack &
Inference \\
\cline{1-1}\cline{3-7}
AutoCCL~\cite{autoccl} &
&
Searches NCCL algorithms, protocols, and channel configurations online &
\revision{1.07--1.32$\times$ training speedup over NCCL} &
Early runtime &
CCL tuning &
Training \\
\cline{1-1}\cline{3-7}
MCCS~\cite{mccs} &
&
Coordinates paths, QoS, and traffic engineering through a provider service &
\revision{Up to 2.4$\times$ algorithm-bandwidth improvement over NCCL} &
Runtime &
Cloud infrastructure &
Training \\
\hline
CoCoNet~\cite{coconet} &
\multirow[c]{4}{=}[-2.8\baselineskip]{Coordi-\allowbreak nation} &
Transforms computation graphs to split, reorder, fuse, and overlap operators &
\revision{Up to 1.68$\times$ training speedup over DDP; 1.77$\times$ inference speedup over Megatron-LM} &
Compile/\allowbreak autotune &
Computation graph &
Both \\
\cline{1-1}\cline{3-7}
Centauri~\cite{centauri} &
&
Partitions communication to expose overlap opportunities across training graphs &
\revision{Up to 1.49$\times$ FSDP+DP training throughput over ZeRO-1/2/3} &
Config./\allowbreak runtime &
Training framework &
Training \\
\cline{1-1}\cline{3-7}
MegaScale-Infer~\cite{megascaleinfer} &
&
Combines module placement and ping-pong pipelining to hide cross-node traffic &
\revision{Up to 1.90$\times$ per-GPU decoding-throughput improvement over TensorRT-LLM} &
Deploy./\allowbreak runtime &
Serving architecture &
Inference \\
\cline{1-1}\cline{3-7}


NanoFlow~\cite{nanoflow} &
&
Uses nano-batches and dependency-aware scheduling to overlap computation, memory access, and communication &
\revision{1.91$\times$ average throughput improvement over TensorRT-LLM} &
Runtime &
Inference runtime &
Inference \\
\hline
\end{tabular}

\vspace{1mm}
\begin{minipage}{0.98\textwidth}
\footnotesize
\textit{Abbreviations:} ``General'' denotes a communication-layer design not tied to
a specific training or inference workflow.
\end{minipage}
\end{table*}

\section{Communication Planning}
\label{sec:comm plan}
Communication planning determines the logical schedule of a collective operation before it
is realized by a concrete communication runtime. Given a collective primitive, message size,
parallelism demand, and cluster topology, the planning stage decides which workers communicate,
what data chunks are exchanged, and in what order the exchanges occur. Recent work follows two
complementary directions. Topology-aware handcrafted planning redesigns collective algorithms using
expert knowledge of network hierarchy, bandwidth asymmetry, and topology geometry. In contrast,
optimization-based schedule synthesis formulates collective design as a search or optimization
problem and automatically generates schedules from topology and workload constraints. Together,
these two directions show the evolution of communication planning from manual topology-aware
algorithm design toward more systematic and scalable schedule generation.

\subsection{Topology-Aware Handcrafted Planning}
Topology-aware handcrafted planning starts from a basic but often overlooked fact: a collective schedule is a projection of logical data movement onto a physical interconnect. Modern GPU clusters are not uniform communication fabrics. GPUs inside a server may communicate through NVLink, NVSwitch, or PCIe, while cross-server traffic must pass through NICs and the data-center network. These links differ in bandwidth, latency, contention behavior, and sharing scope. As a result, a schedule that looks balanced in a logical ring or tree can overload slow inter-node links and underuse fast intra-node links. It can even cause synchronization stragglers and tail latency where some chunks wait for traffic on bottleneck paths to finish.

\revision{BlueConnect~\cite{blueconnect} is designed for hierarchical GPU clusters. It recursively decomposes AR into RS and AG, keeping most reduction on high-bandwidth intra-node links before reduced data traverses slower inter-node links. On 192 GPUs for ResNet-50 training, this design reduced synchronization overhead by 87\% relative to GLOO. Its main advantage is a transparent schedule that can be deployed efficiently when the hierarchy is known and stable.}
\revision{Themis~\cite{themis} extends this principle to clusters whose network dimensions have unequal bandwidths. It schedules chunks in proportion to the capacity of different dimensions, reporting average training-iteration speedups of 1.25--1.49$\times$ across four workloads. The progression from BlueConnect to Themis is therefore from avoiding slow cross-level traffic to balancing the residual traffic across heterogeneous links. Both approaches are most suitable when accurate topology and bandwidth information is available; their performance can be less predictable when those assumptions become stale because of changing placement or contention. This dependence on expert topology reasoning motivates schedule synthesis.}

\subsection{Optimization-Based Schedule Synthesis}

Handcrafted planning shows that collective performance depends strongly on the match between a communication schedule and the underlying topology. However, deriving such schedules manually becomes increasingly difficult as systems scale. A modern training or inference deployment may contain multiple link types, nonuniform bandwidth, different collective primitives, and parallelism groups of different sizes. A schedule optimized for a specific configuration may no longer be suitable when the system or workload changes. Optimization-based schedule synthesis addresses this limitation by making the planning problem explicit: given a collective primitive, a topology, and a performance objective, the system searches for a communication schedule instead of relying on a manually designed algorithm.

Schedule synthesis operates at a finer granularity than selecting a predefined collective algorithm. It treats the collective as a set of chunk-level data movements subject to correctness and resource constraints. For each chunk, the synthesizer determines its communication endpoints, timing, and dependencies. Different choices can implement the same collective primitive but induce very different traffic patterns on the physical network. The goal is therefore to find a schedule that preserves the collective semantics while using available links efficiently and reducing completion time.

\revision{SCCL~\cite{sccl} encodes collective semantics, topology, and ordering decisions as solver constraints and searches for a valid chunk-level schedule. This specialization can outperform fixed library algorithms; for example, its synthesized AG schedules were up to 2.2$\times$ faster than NCCL for small inputs on the evaluated topology. The trade-off is that the search space grows with the number of devices, chunks, and communication steps, making direct synthesis increasingly costly.}

\revision{TE-CCL~\cite{teccl} shifts the formulation toward traffic engineering by treating chunks as multi-commodity flows over capacity-limited links. This makes routing, capacity allocation, and congestion explicit and achieves up to 3.18$\times$ higher algorithm bandwidth than RCCL on its GPU testbed. Relative to SCCL, this view is particularly useful when link capacities and shared bottlenecks dominate the schedule. Compared with handcrafted strategies, synthesis offers a more systematic way to explore efficient schedules across changing topologies and workload traces. However, its search space can grow explosively as the number of devices and the scale of the communication problem increase.}

\revision{Later work such as SyCCL~\cite{syccl} mitigates this search-space explosion by exploiting symmetry in collective demands and cluster topologies to prune the search space in advance. This restriction creates a trade-off between the quality of the generated schedules and synthesis time. Such synthesizers are limited to schedule-level optimization: in very large clusters, final performance still relies on experts co-optimizing the schedule with runtime execution and network tuning.}

\section{Communication Execution and Adaptation}
\label{sec:comm exec}
A schedule must be implemented by a communication runtime and mapped to the resources of a real cluster. Recent work therefore focuses on two aspects: programmable execution, which provides abstractions for building custom collective programs, and runtime adaptation, which adjusts communication behavior according to hardware topology and network conditions.
\subsection{Programmable Collective Execution}
A planned schedule can improve performance only when the communication runtime can express it and execute it efficiently. Planning specifies logical data movement, including chunk transfers, communication endpoints, and ordering constraints, but execution must realize these decisions through GPU kernels, buffers, synchronization, and transport operations. This creates a practical gap between schedule design and communication libraries. Conventional CCL interfaces usually expose collectives as fixed calls, such as AR or AG, and hide the internal schedule from applications. While this abstraction simplifies programming, it makes schedules tailored to a specific topology or workload difficult to deploy.


\revision{To provide both high performance and programmability for AI applications, MSCCL++~\cite{mscclpp} introduces a layered GPU communication stack. It exposes primitive communication functions through straightforward interfaces, enabling GPU experts to implement fine-grained optimizations while providing higher-level interfaces for rapid workload-specific customization. This layered design reduces development and optimization effort and enables faster adoption of emerging hardware capabilities. It achieved 1.7$\times$ geometric-mean collective speedup and 1.2$\times$ geometric-mean inference speedup against the evaluated baselines. It is particularly suitable for GPU communication experts and system developers who need to tailor communication to specific inference workloads. Although the Collective API lowers adoption effort, it is still limited to inference workloads. Its portability also depends on the availability of a Primitive API implementation for the target hardware and interconnect.}

\subsection{Adaptive Collective Execution}
Programmability makes customized collective schedules easier to express and deploy, but execution performance still depends on the deployed environment. The same collective can behave differently as message sizes, runtime parameters, workload behavior, and network load change. As a result, a fixed execution choice may work well in one setting but become inefficient in another. Adaptive collective execution addresses this problem by adjusting how a collective is carried out according to runtime observations.

\revision{AutoCCL~\cite{autoccl} adapts within the application runtime. It profiles algorithms, protocols, and channel configurations during early training iterations, then reuses the best configuration for recurring collectives. It improved end-to-end per-iteration training time by 1.07--1.32$\times$ over NCCL. Its effectiveness is greatest for long-running training jobs with recurring communication patterns and substantial communication volume, where application-level visibility of message behavior and computation interference allows the early profiling cost to be amortized over many iterations. It is less suitable when communication patterns vary frequently, since the profiling process may not converge to a stable, near-optimal configuration.}

\revision{In multi-tenant clouds, individual tenants cannot observe the physical topology, link utilization, or communication activities from other tenants, making it difficult to select an efficient collective strategy. MCCS~\cite{mccs} therefore moves collective communication from a tenant-side library to a provider-managed service. The service preserves the collective API while allowing the cloud provider to select and dynamically reconfigure collective strategies using its global view of shared network resources. MCCS achieves up to 2.4$\times$ higher algorithm bandwidth than NCCL. Compared with AutoCCL, MCCS coordinates communication across tenants at the infrastructure layer, whereas AutoCCL tunes configurations locally for an individual training job. AutoCCL is thus better suited to relatively stable environments with recurring communication patterns; under cross-tenant interference, its locally selected configuration may no longer remain optimal.}

\section{Computation-Communication Coordination}
\label{sec:codesin}
Optimizing an individual collective does not necessarily improve an end-to-end LLM
workload. The benefit depends on where communication sits in the execution graph and whether it can
be hidden behind useful computation. This section therefore shifts from collective operations alone
to their interaction with model execution. Training workloads provide repeated computation graphs,
which expose stable overlap windows. Inference workloads are less regular: token routing, small
batches, prefill-decode imbalance, and latency targets make communication part of the online serving
path itself. 
These two settings motivate a split discussion between training-side and inference-side communication.

\subsection{Communication Scheduling in Training Graphs}
Distributed training repeatedly executes the same forward and backward structure over many iterations. Collective operations such as gradient synchronization and parameter gathering appear at stable dependency points in the iteration graph. If these collectives remain on the critical path, they directly increase iteration time and reduce throughput; if they overlap with independent computation, part of their cost can be hidden. Training-side optimization thus focuses on placing collectives within the repeated execution graph so that less communication is exposed on the critical path.

\revision{In hybrid-parallel LLM training, communication from DP, fully sharded data parallelism (FSDP), TP, PP, and EP can all lie on the critical path, often at different points in the training graph. Computation-communication co-design is therefore needed to coordinate these operations and reduce their exposed cost, rather than optimizing each collective in isolation.}

\revision{CoCoNet~\cite{coconet} uses a unified DSL to describe computation and communication together, and generates customized optimized code after applying dependency-preserving transformations. It reported up to 1.68$\times$ speedup for BERT training over PyTorch DDP. For GPT-2 inference, it achieved up to 1.77$\times$ speedup over Megatron-LM. By optimizing kernels automatically across DP, TP, and PP, this design is particularly suitable for iterative training workloads, where stable graph structures can amortize the tuning overhead. Its kernel-level optimizations of TP and PP execution also improve inference performance. However, its fine-grained kernel fusion can miss opportunities for graph-level scheduling; moreover, fusion can even reduce performance for small tensors.}

\revision{Centauri~\cite{centauri} targets hybrid LLM training with DP, FSDP, TP, and PP from a framework-level perspective. It expands the overlap space by partitioning a collective along primitive, topology-aware group, and workload dimensions, so that sub-communications can become ready at different points and can be interleaved with their dependent computation. Implemented in Megatron-LM, Centauri achieves up to 1.49$\times$ higher training throughput than ZeRO-1/2/3 in FSDP+DP configurations. In contrast to CoCoNet, Centauri rearranges partitioned computation and communication at the scheduling level. As the model size scales, FSDP is commonly used to reduce memory pressure. Centauri is well-suited to large-scale FSDP and hybrid-parallel training.}

\subsection{Communication Integration in Inference Pipelines}
Inference provides fewer and less predictable opportunities for hiding communication than training. Request batches, sequence lengths, and token-routing decisions change over time, while autoregressive decoding processes only a limited number of tokens in each iteration. These properties reduce the computation available for conventional overlap and make communication directly visible in time to first token, time per output token, and tail latency. \revision{Recent work therefore addresses inference communication at multiple levels, including pipeline orchestration, fine-grained execution scheduling, and serving architecture design.}


\revision{At the pipeline level, MegaScale-Infer~\cite{megascaleinfer} targets large-scale MoE decoding by disaggregating attention and expert (FFN) modules. Attention and expert nodes can scale independently with tailored parallelism and hardware choices. Its ping-pong pipeline divides requests into micro-batches and alternates their execution between the two modules, allowing cross-module communication to be hidden by computation. It reports up to 1.90$\times$ higher per-GPU decoding throughput than TensorRT-LLM. Its benefit diminishes when attention and expert execution cannot be balanced or cross-module communication remains exposed, as pipeline bubbles can offset the gain.}


\revision{At the execution level, NanoFlow~\cite{nanoflow} explores computation-communication overlap by restructuring LLM inference execution into fine-grained nano-batches. NanoFlow partitions requests into smaller execution units and schedules nano-operations according to their dependencies, exposing concurrency among GPU computation, memory access, and communication. This design exposes overlap opportunities that are difficult to achieve with conventional batch-level scheduling and improves resource utilization under dynamic serving workloads. It reports 1.91$\times$ higher throughput on average than TensorRT-LLM. However, its benefit depends on sufficient workload parallelism and fine-grained scheduling opportunities; for small requests, latency-sensitive workloads, or scenarios where communication cannot be effectively overlapped with independent computation, the additional scheduling complexity provides limited improvement.}

\revision{At the serving architecture level, growing model sizes and request volumes make the serving architecture itself an important communication consideration. Prefill processes an entire prompt and is computation-intensive, whereas autoregressive decoding generates one token at a time and is dominated by key-value (KV) cache access and memory bandwidth. Prefill--decode (PD) disaggregation separates the two phases so that each can use tailored resources and parallelism, avoiding the compromise of a shared GPU group. Its main challenge is the additional network communication required to transfer the KV cache and intermediate states between the two groups; this cost can directly affect TTFT and becomes more pronounced for long-context requests~\cite{distserve}.}

\section{Open Challenges and Future Opportunities}
\label{sec:challenges}

Collective communication optimization is moving beyond isolated primitive tuning toward broader system-level challenges. 
As LLMs continue to scale and their services become more widely used, collective communication must support increasingly diverse and demanding workloads. We provide several critical directions for future research on collective communication.

\noindent\textbf{Adaptive Collectives under Dynamic Workloads.}
Traditional collective communication assumes static and predictable execution patterns, which no longer hold in modern and future LLM systems. 
During large-scale training, shared multi-tenant cloud environments exhibit frequent network jitter and dynamic bandwidth fluctuations over long-running jobs. In inference pipelines, unpredictable user request arrivals trigger bursty traffic, fluctuating batch sizes, and highly dynamic token routing patterns in MoE models. 
Static communication algorithms or one-time profiling fail to sustain optimal performance under such high variance. 
\revision{At the execution layer, collective runtimes need a lightweight feedback loop that observes message sizes, link utilization, and computation interference. At safe collective boundaries, it can adapt the logical topology, chunk size, and routing path to workload changes. The key is to keep monitoring and reconfiguration overhead low while preventing unstable configuration oscillations.}

\noindent\textbf{Unified Communication Library.}
Current LLM frameworks rely on many specialized communication libraries, such as NCCL for general
collectives, DeepEP~\cite{deepep} for MoE expert dispatch, and
MegaScale-Infer~\cite{megascaleinfer} for disaggregated inference communication. 
However, these libraries often operate as isolated silos, leading to duplicated resource management where each co-running library independently allocates redundant monitoring frameworks, network contexts, and memory buffers. 
More importantly, because these libraries are mutually oblivious to each other's traffic patterns, they inevitably compete blindly for shared physical bandwidth without global coordination. This uncoordinated competition triggers severe network congestion and degrades tail latency. 
\revision{A unified communication library should separate collective semantics from topology-specific scheduling and execution. Under a common interface, it can select or generate specialized schedules for different topologies and workloads. The challenge is to retain such specialization and performance isolation when multiple collectives share the same communication resources, rather than forcing them into a single generic implementation.}

\noindent\textbf{Portable Communication across Heterogeneous Hardware.}
Existing communication libraries are deeply coupled with specific accelerator architectures (e.g., GPU for NVIDIA, NPU for Huawei), customized network interfaces, and dedicated interconnects (e.g., NVLink/NVSwitch for NVIDIA, HCCS/UnifiedBus for Huawei). As the hardware ecosystem diversifies with a mixture of different GPUs, ASICs, SmartNICs, and heterogeneous network fabrics, porting and manually retuning these monolithic libraries becomes prohibitively expensive and even infeasible. Current hardware-agnostic abstractions often sacrifice peak hardware performance for generality, failing to exploit specialized architectural features. \revision{At the execution layer,} future research must bridge this gap by designing portable yet highly performant communication abstractions. The key lies in decoupled system architectures that can automatically compile high-level communication intent into hardware-specific optimizations. Such frameworks should leverage unified intermediate representations (IRs) to achieve cross-hardware portability without compromising the ability to utilize fine-grained, device-level capabilities.

\noindent\textbf{Compression-Aware Collective Communication.}
As communication bottlenecks worsen relative to compute scaling, reducing transferred data volume via lossy quantization, sparsification, or algorithmic compression has emerged as a compelling strategy. 
However, modern communication primitives are largely compression-blind, treating transferred tensors as opaque bitstreams. Integrating data reduction techniques into collective communication introduces non-trivial system-level trade-offs, where serialization and encoding overheads on accelerators often offset the network savings, and error accumulation can severely degrade model accuracy. Future communication systems should co-design data representations and collective execution semantics \revision{at the execution and adaptation layer}. This requires building compression-aware collective primitives that can adaptively select compression ratios and schemes based on real-time network conditions and hardware traits. Furthermore, future systems should explore in-network or on-the-fly compression to overlap encoding latency with data transmission.


\noindent\textbf{Communication Offloading.}
A major architectural limitation of many dominant communication libraries is their heavy reliance on GPU computing resources (such as GPU Streaming Multiprocessors and device memory bandwidth) to drive network progress and orchestrate collectives. This dependency causes severe resource contention and hardware interference between communication and computation workloads. This issue becomes particularly critical in inference and serving systems where communication sits directly on the user-facing latency critical path, choking GPU utilization. 
\revision{Current in-network aggregation mechanisms offload collective reduction to network devices. Beyond this mechanism, offloading parts of communication processing from GPUs to CPUs or dedicated copy engines is already practical in many systems. Future systems can further leverage DPUs and SmartNICs for broader control and execution tasks. The remaining challenge is to co-design computation and communication so that offloading decisions improve end-to-end performance under limited physical resources.}

\noindent\textbf{Fault Tolerance and Reliability.}
While existing collective communication research predominantly focuses on maximizing raw throughput, large-scale LLM deployment demands robust system reliability and high availability. 
In clusters consisting of tens of thousands of nodes, hardware failures and silent data corruption become increasingly likely at scale.
Under current rigid collective paradigms, a single slow node (straggler) or a failed rank can easily stall the entire collective operation, leading to catastrophic training restarts or unacceptable SLO violations during inference serving. 
\revision{Fault tolerance primarily exposes a gap at the execution and runtime-adaptation layer. Upon a failure signal or evidence of a persistent straggler, the collective runtime must identify the affected communication context and repair only the necessary communicator links. Mnemosyne~\cite{mnemosyne} couples framework-level just-in-time recovery with a flexible CCL that dynamically adjusts affected links, avoiding global communication reinitialization. However, local communication recovery must be coordinated with framework-side state restoration and task replay to preserve model and optimizer state. The remaining cross-layer challenge is to combine failure sensing and local collective repair with framework recovery, so that a local failure does not trigger an unnecessary global restart or stall unaffected ranks.}

\section{Conclusion}
\label{sec:conclusion}
Collective communication optimization is undergoing a paradigm shift, evolving from accelerating isolated communication primitives to addressing a complex, cross-layer system challenge in large-scale LLM training and serving. 
To capture this evolution, this survey provides a holistic, full-stack perspective on collective communication within modern distributed ecosystems. Specifically, we systematically categorize recent literature into three interconnected pillars: communication planning, runtime execution and adaptation, and computation-communication coordination. Furthermore, a comprehensive taxonomy and comparison of representative frameworks are provided to highlight the trade-offs of existing solutions. 
Finally, we map out critical open challenges and future research opportunities. 
We hope this survey serves as a valuable roadmap, inspiring next-generation collective communication optimizations in this LLM era.

\section*{Acknowledgments}
This work is supported by the National Key Research and Development Program of China (2024YFB4505604), the Open Research Fund of State Key Laboratory of Internet Architecture (HLW2025ZD01), the Shandong Provincial Natural Science Foundation of China (ZR2024LZH011), the Fundamental Research Funds for the Central Universities, and Beihang-Infrawaves Joint Lab.



\bibliographystyle{IEEEtran}
\bibliography{refs}


\section*{Biography Section}
 




\begin{IEEEbiographynophoto}
{Xuebin Song} (soybean@buaa.edu.cn) 
received his B.S. degree from the School of Software, Beihang University, Beijing, China, and
is currently pursuing his Master's degree with the School of Software, Beihang University, Beijing, China. His current research interests include LLM training and collective communication optimization.
\end{IEEEbiographynophoto}

\begin{IEEEbiographynophoto}
{Menghao Zhang} (Member, IEEE) (zhangmenghao@buaa.edu.cn)
received his B.S. degree and Ph.D. degree in computer science from Tsinghua University, Beijing, China. He is currently an Associate Professor with the School of Software, Beihang University, Beijing, China. His research interests include high-performance networks, networked systems, LLM systems, and network security.
\end{IEEEbiographynophoto}

\begin{IEEEbiographynophoto}
{Yuezheng Liu} (liuyuezheng@buaa.edu.cn)
received his B.S. degree from the School of Computer Science and Engineering, Beihang University, Beijing, China, and
is currently pursuing a Master's degree with the School of Computer Science and Engineering, Beihang University, Beijing, China. His current research interests include LLM networked systems.
\end{IEEEbiographynophoto}

\begin{IEEEbiographynophoto}
{Jinyi Xia} (xiajinyi@buaa.edu.cn)
received his B.S. degree in information security from Beijing University of Posts and Telecommunications, Beijing, China, and is currently pursuing a Ph.D. degree in software engineering from Beihang University, Beijing, China. His research interests include networked systems and LLM systems.
\end{IEEEbiographynophoto}

\begin{IEEEbiographynophoto}
{Shucan Yang} (landmvrks@163.com) 
received his B.S. degree in School of Computer Science, Beijing University of Posts and Telecommunications, Beijing, China, and is currently pursuing a Ph.D. degree in software engineering from Beihang University, Beijing, China. His current research interests include networked systems and LLM systems.
\end{IEEEbiographynophoto}

\begin{IEEEbiographynophoto}
{Xiaohe Hu} (huxiaohe@infrawaves.com)
received the B.S. and Ph.D. degrees from the Department of Automation, Tsinghua University, Beijing, China. He is currently the CEO of the Infrawaves, Beijing, China. His main research interests include AI infra and LLM systems.
\end{IEEEbiographynophoto}

\begin{IEEEbiographynophoto}
{Chunming Hu} (Member, IEEE) (hucm@buaa.edu.cn)
received the B.S. and Ph.D. degree in computer science from Beihang University, Beijing, China. He is a professor with the School of Software, Beihang University, Beijing, China. His research interests include distributed systems, system virtualization, data management and processing systems.
\end{IEEEbiographynophoto}

\begin{IEEEbiographynophoto}
{Mingwei Xu} (xumw@tsinghua.edu.cn)
received the B.S. and Ph.D. degrees from Tsinghua University, Beijing, China. He
is a full professor with the Department of Computer Science, Tsinghua University. His research interests include computer network architecture, high-speed router architecture, and network security.
\end{IEEEbiographynophoto}

\vfill

\end{document}